\documentclass[preprintnumbers,10pt,nofootinbib]{revtex4}
\pdfoutput=1

\usepackage{amsmath,latexsym,amssymb,amsfonts}
\usepackage[pdftex]{color,graphicx}
\usepackage{bm}

\begin{document}


\title{Phantom of the Hartle-Hawking instanton: connecting inflation with dark energy}

\author{\textsc{Pisin Chen}$^{1,2}$\footnote{{\tt pisinchen@phys.ntu.edu.tw}},\;\; \textsc{Taotao Qiu}$^{3}$\footnote{{\tt qiutt@mail.ccnu.edu.cn}}\;\; and \; \textsc{Dong-han Yeom}$^{1}$\footnote{{\tt innocent.yeom@gmail.com}}}

\affiliation{$^{1}$\small{Leung Center for Cosmology and Particle Astrophysics, National Taiwan University, Taipei 10617, Taiwan}\\
$^{2}$\small{Kavli Institute for Particle Astrophysics and Cosmology, SLAC National Accelerator Laboratory, Stanford University, CA 94305, USA}\\
$^{3}$\small{Institute of Astrophysics, Central China Normal University, Wuhan 430079, People's Republic of China}
}

\begin{abstract}
If the Hartle-Hawking wave function is the correct boundary condition of our universe, the history of our universe will be well approximated by an instanton. Although this instanton should be classicalized at infinity, as long as we are observing a process of each history, we may detect a non-classicalized part of field combinations. When we apply it to a dark energy model, this non-classicalized part of fields can be well embedded to a quintessence and a phantom model, i.e., a quintom model. Because of the property of complexified instantons, the phantomness will be naturally free from a big rip singularity. This phantomness does not cause perturbative instabilities, as it is an effect \textit{emergent} from the entire wave function. Our work may thus provide a theoretical basis for the quintom models, whose equation of state (EoS) can cross the cosmological constant boundary (CCB) phenomenologically.
\end{abstract}

\maketitle

\newpage

\tableofcontents


\section{Introduction}

One of the crucial tasks of quantum gravity is to understand the singularities in general relativity. When we consider the initial singularity of our universe \cite{Hawking:1969sw}, the problem is related to various issues of physical cosmology, e.g., the origin of the emergence of time, the initial condition of inflation, the typicalness of our universe, etc. To deal with these issues, the traditional approach is to investigate the canonical quantization and to study the wave function of the universe \cite{DeWitt:1967yk}.

After invoking the canonical quantization that includes the metric, what we eventually obtain is the master wave equation, the Wheeler-DeWitt equation. This equation is a partial differential equation and hence it requires boundary conditions. We do not know what should be the correct boundary condition, but perhaps the ground state of the universe can be a reasonable choice. Hartle and Hawking (HH) \cite{Hartle:1983ai} suggested that the Euclidean path integral provides a good analog of the ground state wave function. For cosmological applications, the $O(4)$ symmetric metric ansatz would be a good simplification; and the Euclidean path integral can be approximated by the steepest-descent approximation, or by sum-over instantons. When we consider the Euclidean instantons, we need to complexify the time and hence every fields should be complexified by analyticity \cite{Halliwell:1989dy}. However, as long as the field is complex-valued, classical properties can never be restored in terms of equations of motion. Therefore, after the Wick rotation, the reality of the metric and the matter field is required: this is the \textit{classicality}\footnote{By `classicality' we mean that a universe is classical, where the universe is originated from the wave function and the wave function itself is not classical. This is different from another notion of `classicality' in the literature of inflationary physics; in this context, people consider a classicalization of quantum fluctuations. In this paper, our physical object and interest are different from the latter issue (classicalization of quantum fluctuations).} condition \cite{Hartle:2007gi,Hartle:2008ng}.

Already some techniques have been investigated to calculate HH instantons and to estimate the probability distribution of each initial conditions \cite{Hartle:2007gi,Hartle:2008ng,Hwang:2011mp}. Typical expectations of the HH wave function are as follows: (1) it provides slow-roll inflation to obtain a classical history and (2) it does not prefer large number of $e$-foldings. The former is useful, but the latter is not a good news for inflationary cosmology \cite{Guth:1980zm}. However, if one considers more sophisticated models, this difficulty can be resolved. Note that when our universe begins, all field should be realized and satisfy the classicality condition. We envision that there exists some fields that in the early universe: an inflaton (or inflatons) that induces inflation, heavy mass fields and light mass fields compared with the inflaton. Regarding such a setting, the followings should be noticed.
\begin{itemize}
\item[--] If the mass scales of the fields are similar, then it is reasonable that these fields are equally excited at the same time. This may be related to the assisted inflation of multi-fields that can help to prefer large $e$-foldings \cite{Hwang:2012bd}.
\item[--] If there is a much heavier mass field (or fields), then in order to classicalize the heavier mass field, the lighter field should be excited \cite{Hwang:2014vba}. This excited lighter field can in principle be the inflaton field, which may further explain the preference of sufficiently large $e$-foldings.
\item[--] Some modifications of the gravity sector in the early universe may help to prefer large $e$-foldings \cite{Sasaki:2013nka}.
\end{itemize}
While this is not yet settled, it is fair to say that the HH wave function remains a reasonable theoretical basis for our inflationary universe \cite{Hwang:2013nja}.

If so, then the natural next question is, what will happen to the much lighter fields compared with the inflaton? Of course, at once they exist from the beginning, then these light fields should be regarded as a part of instanton. At the first glimpse, it is natural to assume that these light fields should be classicalized, too. However, if a field is decoupled from our phenomenological fields (standard model particles) and the amount of energy of this field is much smaller than that of the inflaton field, then even though the field is not classicalized, there is no way to distinguish the light field during and after the primordial inflation. As time goes on, however, the super slow-rolling and non-classicalized field can leave some distinguishable effects in the universe around the dark energy dominated era. This is a kind of `residue' from the quantum gravity. Then can we see these effects in this universe? (Regarding this topic, for extended calculations, see \cite{Chen}.)

Motivated by this scenario, in this paper we study the properties of a field that has \textit{negligible amount of energy} compared to the inflaton, which is \textit{super slow-rolling} and \textit{non-classicalized}. By non-classicalized, we mean that the scalar field is not entirely realized from complex values (following the notion of classicality in \cite{Hartle:2007gi,Hartle:2008ng}). Although this field has the form of a quintessence field, however, due to its non-classicalicity, some part of this field will also possess the phantom behavior. Therefore, effects of the non-classicalized field can be very well-embedded in a quintessence + phantom dark energy model, i.e., the quintom model \cite{Feng:2004ad}. This quintum model is known to be useful to investigate late time cosmology, especially in order to explain the crossing phenomenon of the dark energy equation of state over the cosmological constant boundary. Now the question is this: if there remain effects from a non-classicalized field as a quintum model, then what will be the signatures to our late time universe? This is the task of this paper.

This paper is organized as follows. In SEC.~\ref{sec:hh}, we briefly summarize previous results on the HH wave function. In SEC.~\ref{sec:ltc}, we discuss the behavior of the non-classicalized field that is indeed a quintom model; we also discuss the physical implications of this model. In SEC.~\ref{sec:int}, we discuss further interpretational issues, and finally, in SEC.~\ref{sec:con}, we summarize this paper and discuss future issues that should be further investigated.

\section{\label{sec:hh}Hartle-Hawking wave function for two scalar fields}

\subsection{Basic formalism and classicality}

The ground state wave function by Hartle and Hawking \cite{Hartle:1983ai} is defined as the Euclidean path integral for a compact $3$-dimensional manifold $\Sigma$ as a functional of the $3$-metric $h_{\mu\nu}$ and the field value $\chi$ by
\begin{eqnarray}
\Psi[h_{\mu\nu}, \chi] = \int_{\mathcal{M}} \mathcal{D}g_{\mu\nu} \mathcal{D} \phi \; e^{-S_{\text{E}}[g_{\mu\nu}, \phi]},
\label{general_wave_function}
\end{eqnarray}
where the $4$-metric $g_{\mu\nu}$ and the field $\phi$ (for multi-field case, include all fields) take the value $h_{\mu\nu}$ and $\chi$ on $\Sigma = \partial \mathcal{M}$, where $\mathcal{M}$ is a compact $4$-dimensional Euclidean manifold. We integrate over all $\mathcal{M}$ that have $\Sigma$ as their only boundary.

In this paper, we investigate Einstein gravity with two minimally coupled scalar fields $\phi_{1,2}$ (we choose the units $c=G= \hbar = 1$):
\begin{eqnarray}
S_{\text{E}} = - \int dx^{4} \sqrt{+g} \left( \frac{1}{16\pi} R - \sum_{i=1,2} \frac{1}{2} (\nabla \phi_{i})^{2} - V(\phi_{1},\phi_{2}) \right).
\end{eqnarray}
For the purpose of demonstrating qualitative properties, here we invoke a simple quadratic potential with mass $m_{1}$ and $m_{2}$:
\begin{eqnarray}
V(\phi_{1},\phi_{2}) = V_{0} + \frac{1}{2} m_{1}^{2} \phi_{1}^{2} + \frac{1}{2} m_{2}^{2} \phi_{2}^{2}.
\end{eqnarray}
What we want to attain are the following conditions:
\begin{itemize}
\item[1.] $V_{0}$ is much smaller than $m_{1}^{2}$: $V_{0}/m_{1}^{2} \ll 1$. Therefore, during the inflationary era, we can ignore $V_{0}$.
\item[2.] $\phi_{2}$ satisfies over-damped conditions even with $V_{0}$: $m_{2}^{2}/V_{0} < 6\pi$ (or, $m_{2}/\tilde{H} < 3/2$, where $ \tilde{H}^{2} = 8\pi V_{0}/3$). Therefore, after the inflation era, $\phi_{2}$ still satisfies the over-damped condition.
\end{itemize}

\paragraph{Minisuperspace model}

\begin{figure}
\begin{center}
\includegraphics[scale=0.4]{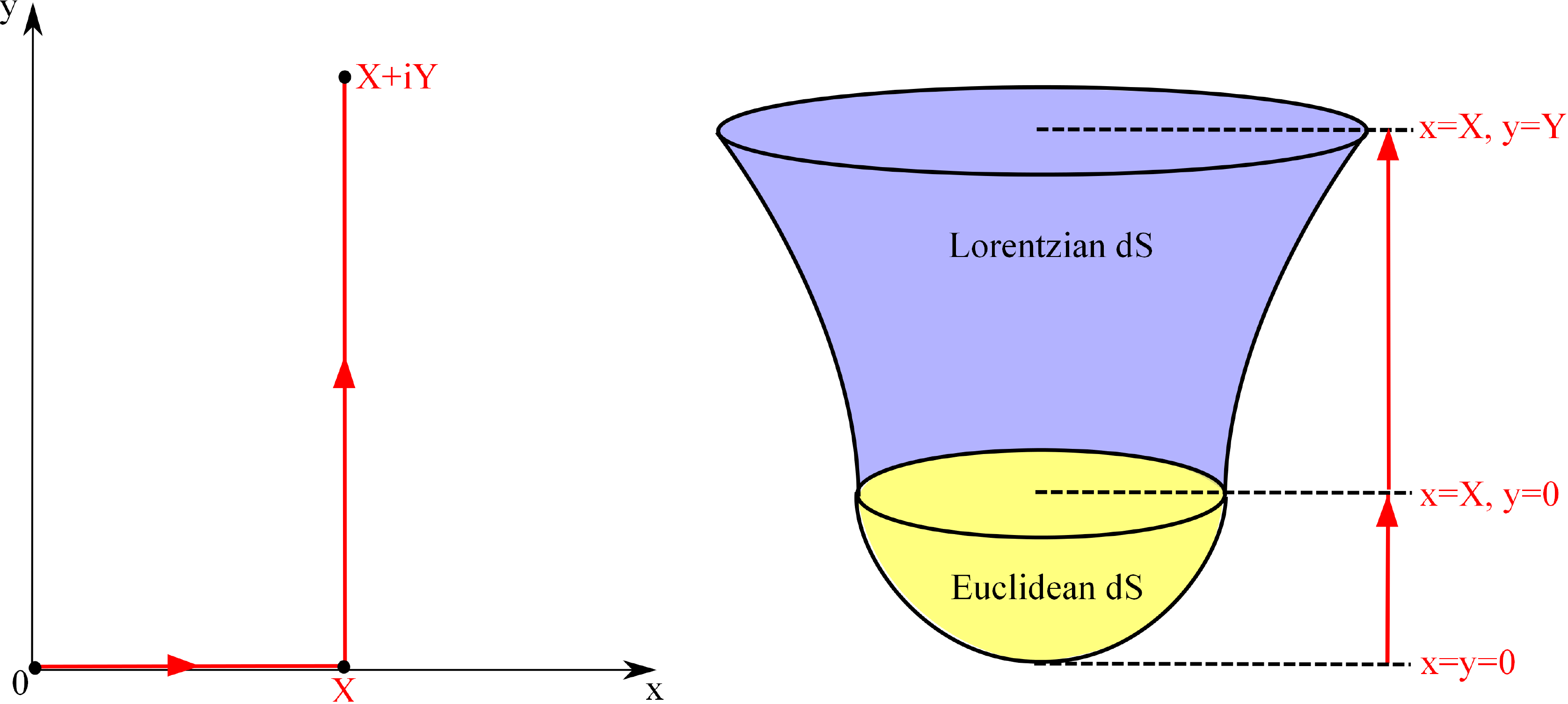}
\caption{\label{fig:path}Left: An instanton solution is defined on the complex plane $\tau = x + iy$. Right: By choosing a contour $\mathcal{C}$ (red arrows) we can draw a combination of the Euclidean and the Lorentzian manifolds. If we choose a proper initial condition and a proper turning time $X$, we can satisfy the classicality condition at large $Y$.}
\end{center}
\end{figure}

We impose the minisuperspace model following the $O(4)$ symmetric metric ansatz
\begin{eqnarray}
ds_{\mathrm{E}}^{2} = \frac{d\tau^{2} + a^{2}(\tau) d\Omega_{3}^{2}}{m_{1}^{2}}.
\end{eqnarray}
From this choice of metric, it is convenient to redefine
\begin{eqnarray}
\mu \equiv \frac{m_{2}}{m_{1}}.
\end{eqnarray}
The HH wave function is now
\begin{eqnarray}
\Psi[b,\chi_{1},\chi_{2}] = \int_{\mathcal{C}} \mathcal{D}a \mathcal{D}\phi_{1} \mathcal{D}\phi_{2} \; e^{-S_{\text{E}}[a,\phi_{1},\phi_{2}]},
\label{minisuperspace_wave_function}
\end{eqnarray}
where the action is reduced by (here, we ignored the $V_{0}$ term)
\begin{eqnarray}
S_{\mathrm{E}} = \frac{2 \pi^{2}}{m_{1}^{2}} \int d\tau \left[- \frac{3}{8\pi} \left(aa^{\prime2} + a \right) + a^{3} \left\{ 1 + \frac{1}{2} \left(\phi_{1}^{\prime2} +\phi_{2}^{\prime2} + \phi_{1}^{2} + \mu^{2} \phi_{2}^{2} \right) \right\} \right].
\end{eqnarray}
Even though $\mu \ll 1$, we explicitly retain this term to study the behavior of the field $\phi_{2}$.
Along the contour $\mathcal{C}$, the metric $a$ starts from zero, which will be interpreted as the South Pole; along this contour, it grows to the boundary value $b$ in the Lorentzian regime where $\phi_{i}$ takes the value $\chi_{i}$ (FIG.~\ref{fig:path}).

\paragraph{Steepest-descent approximation}

To approximately estimate the path-integral, we use the steepest-descent approximation. We approximate the wave function by summing over on-shell histories, the \emph{instantons}, that satisfy the same boundary conditions \cite{Hartle:1983ai}. For such an on-shell history $p$, the HH wave function is approximated by
\begin{eqnarray}
\Psi[b,\chi_{1},\chi_{2}] \simeq \sum_{p} e^{-S_{\text{E}}^{p}}.
\end{eqnarray}

Note that the on-shell condition is to satisfy the following equations of motion:
\begin{eqnarray}
0 &=& a^{\prime\prime} + \frac{8\pi}{3} a \left(\phi_{1}^{\prime2} + \phi_{2}^{\prime2} + \frac{1}{2} \left( \phi_{1}^{2} + \mu^{2}\phi_{2}^{2} \right) \right),\\
0 &=& \phi_{1}^{\prime\prime} + 3 \frac{a^\prime}{a} \phi_{1}^\prime - \phi_{1},\\
0 &=& \phi_{2}^{\prime\prime} + 3 \frac{a^\prime}{a} \phi_{2}^\prime - \mu^{2} \phi_{2},
\label{equations_of_motion}
\end{eqnarray}
where $^\prime$ denotes a derivative with respect to $\tau$.

\paragraph{Classicality condition}

Since our universe follows the Lorentizian signature, a time contour in the path integral (Eq.~\eqref{minisuperspace_wave_function}) should connect from Euclidean to Lorentzian manifold. The contour of $\tau$ is defined on the complex plane (left of FIG.~\ref{fig:path}). The field values at the boundary of the scale factor $b$ and scalar fields $\chi_{i}$ should be real numbers. However, these metric and scalar fields are naturally complexified along the complex time contour. We are interested in the condition of the endpoint ($b$ and $\chi_{i}$). By using the analyticity, we can choose a contour $\tau = x + i y$ for $0 \leq x \leq X$ and $0 \leq y \leq Y$ (right of FIG.~\ref{fig:path}) that connects from $\tau = 0$ to the endpoint. This contour connects from $\tau = 0$ to the turning point at $\tau = X$ through the Euclidean time; then, one can Wick-rotate to the Lorentzian time until the boundary at $\tau = X + i Y$.

If the action along a given history is complex-valued and if the real part and the imaginary part of the action rapidly vary up to the variation of canonical variables, then the Hamilton-Jacobi equation is not satisfied and hence the history is no more classical. On the other hand, if the real part of the Euclidean action varies slowly compared to the imaginary part, then the Hamilton-Jacobi equation (the classical equation of motion) will be approximately satisfied. According to \cite{Hartle:2008ng}, this is called the classicality condition:
\begin{eqnarray}
\left|\nabla_{A} \mathrm{\;Re\;} S_{\mathrm{E}}[b,\chi_{1},\chi_{2}]\right| \ll \left|\nabla_{A} \mathrm{\;Im\;} S_{\mathrm{E}}[b,\chi_{1},\chi_{2}]\right|,
\label{classicality_condition}
\end{eqnarray}
where $A=b,\chi_{i}$. In practice, the classicality condition can be presented by
\begin{eqnarray}
\frac{\left| \mathrm{Im\;} a \right|}{\left| \mathrm{Re\;} a \right|} \ll 1, \;\;\; \frac{\left| \mathrm{Im\;} \phi_{i} \right|}{\left| \mathrm{Re\;} \phi_{i} \right|} \ll 1
\end{eqnarray}
for all $i$'s as $t$ increases, and hence correspond to the reality at the endpoint \cite{Hwang:2014vba}.

When the classicality condition is satisfied, we can interpret that the instanton generates a universe along the time direction. For a classical universe, one can approximate the probability of the Wheeler-DeWitt wave function by
\begin{eqnarray}
P[b,\chi_{1},\chi_{2}] \varpropto \left|\Psi[b,\chi_{1},\chi_{2}]\right|^{2} \simeq e^{-2 \mathrm{\;Re} S_{\mathrm{E}}[b,\chi_{1},\chi_{2}]}.
\label{no-boundary_measure}
\end{eqnarray}

\paragraph{Initial conditions}

The boundary condition at the South Pole comes from the regularity condition,
\begin{eqnarray}
a(\tau = 0) = 0,\qquad
a^\prime(\tau = 0) = 1,\qquad
\phi_{i}^\prime(\tau = 0) = 0.
\end{eqnarray}
At the end endpoint, we impose the following conditions where $b$ and $\chi_{i}$ are real values:
\begin{eqnarray}
a(\tau = X + i Y) = b,\qquad
\phi_{i}(\tau = X + i Y) = \chi_{i}.
\end{eqnarray}
At the turning time, because of the analyticity, we impose the Cauchy-Riemann condition:
\begin{eqnarray}
\frac{\partial a}{\partial x}(\tau = X) = \frac{\partial a}{i \partial y}(\tau = X),\qquad
\frac{\partial \phi_{i}}{\partial x}(\tau = X) = \frac{\partial \phi_{i}}{i \partial y}(\tau = X).
\end{eqnarray}

This system is constructed by second order differential equations of three complex-valued functions: $a$ and $\phi_{i}$. We have eight boundary conditions at $\tau = 0$ and three conditions at the endpoint. We solve this problem by choosing a scalar field value at $\tau = 0$,
\begin{eqnarray}
\phi_{i}(\tau = 0) \equiv \phi_{i}(0) = |\phi_{i}(0)| e^{i \theta_{i}},
\end{eqnarray}
where $|\phi_{i}(0)|$ and $\theta_{i}$ are real. One can solve this initial value problem to calculate time evolutions of $a$ and $\phi_{i}$ from $\tau = 0$. For a given $|\phi_{i}(0)|$, in order to satisfy classicality conditions, one needs to tune $X$ and $\theta_{i}$.

\subsection{Summary of previous results and motivations}

\paragraph{Applications to primordial inflation}

These conclusions are already proven by previous authors:
\begin{itemize}
\item[--] For a single field inflaton with $V = V_{0} + (1/2)m^{2} \phi^{2}$, if $m^{2} / V_{0} < 6\pi$ and hence if the potential is in the slow-roll regime, then the probability distribution is consistent with that of the quantum field theory in de Sitter space \cite{Hwang:2012mf}.
\item[--] On the other hand, if $m^{2} / V_{0} > 6\pi$, then $\phi$ cannot be classicalized around $\phi = 0$. This was proven analytically as well as numerically in \cite{Hartle:2007gi,Hartle:2008ng}.
\item[--] As a simple extension, if there are two fields $\phi_{1}$ and $\phi_{2}$ with $V = (1/2)m_{1}^{2} \phi_{1}^{2} + (1/2)m_{2}^{2} \phi_{2}^{2}$ and $m_{1}/m_{2} \gg 1$ (hence, $\phi_{2}$ direction is a slow-rolling direction), then to classicalize the heavy mass direction $\phi_{1}$ around $\phi_{1} = 0$, we must require the condition \cite{Hwang:2014vba}
\begin{eqnarray}
\frac{m_{1}^{2}}{(1/2) m_{2}^{2} \phi_{2}^{2}} < 6\pi.
\end{eqnarray}
This in turn requires $\phi_{2} \gtrsim (m_{1}/m_{2})$ to classicalize both fields (and, this initial condition is the most probable one as well, see details in \cite{Hwang:2014vba}).
\end{itemize}
In the early universe, there may exist various fields. To classicalize heavy fields, some slow-rolling fields need to be excited and these excited slow-rolling fields can be the origin of inflation.

\paragraph{Motivations: what about a slower direction?}

If the inflaton field is excited, inflation is turned on, and as the inflaton decays, matters and structures will be formed. However, what will happen if there was a much slower direction than the inflaton field? Let us call this field a \textit{quintessence}.

If this quintessence is decoupled from the other matter fields and its direction rolls much more slowly than the inflation itself, then even though the field is not classicalized, it would not induce any observable effect. Hence, \textit{even though the quintessence field is not entirely classicalized, during and post inflation, it renders no observable impact.}

However, at late times after radiation and matter dominant eras, such quintessence field may in principle exhibit some physical imprints. Then what will be the signatures of the non-classicalized quintessence to our late time universe? This physics should be connected to physics of the dark energy, which is the task of this paper.

\section{\label{sec:ltc}Physics of non-classicalized field: quintessence and/or phantom}

We explicitly write the relevant quantities as $a = a_{r} + i a_{i}$, $\phi_{1} = \phi_{1r} + i \phi_{1i}$, and $\phi_{2} = \phi_{2r} + i \phi_{2i}$. Let us assume that $m_{1} \gg m_{2}$, where $\phi_{1}$ is the \textit{inflaton field} and $\phi_{2}$ is the \textit{quintessence field}. In addition, let us further assume that $a$ and $\phi_{1}$ are almost completely classicalized, while $\phi_{2}$ is not. In other words, as $t \rightarrow Y$ (where $t = Y$ is almost the end point of inflation),
\begin{eqnarray}
\frac{\left| a_{i} \right|}{\left| a_{r} \right|} \rightarrow 0, \;\;\; \frac{\left| \phi_{1i} \right|}{\left| \phi_{1r} \right|} \rightarrow 0, \;\;\; \frac{\left| \phi_{2i} \right|}{\left| \phi_{2r} \right|} \simeq \mathcal{O}\left(1\right).
\end{eqnarray}
In addition, we further assume that around the turning time $\tau = X$,
\begin{eqnarray}
\frac{\left| \Omega_{\phi_{2}} \right|}{\left| \Omega_{\phi_{1}} \right|} \ll 1,
\end{eqnarray}
so that the contribution from $\phi_{2}$ does not affect inflationary physics (hence, when the universe is created, the probability is mainly determined by $\phi_{1}$ and does not sensitively depend on $\phi_{2}$).

Then after the primordial inflation and matter/radiation dominated era, there will be an era dominated by the quintessence field.

\begin{figure}
\begin{center}
\includegraphics[scale=0.23]{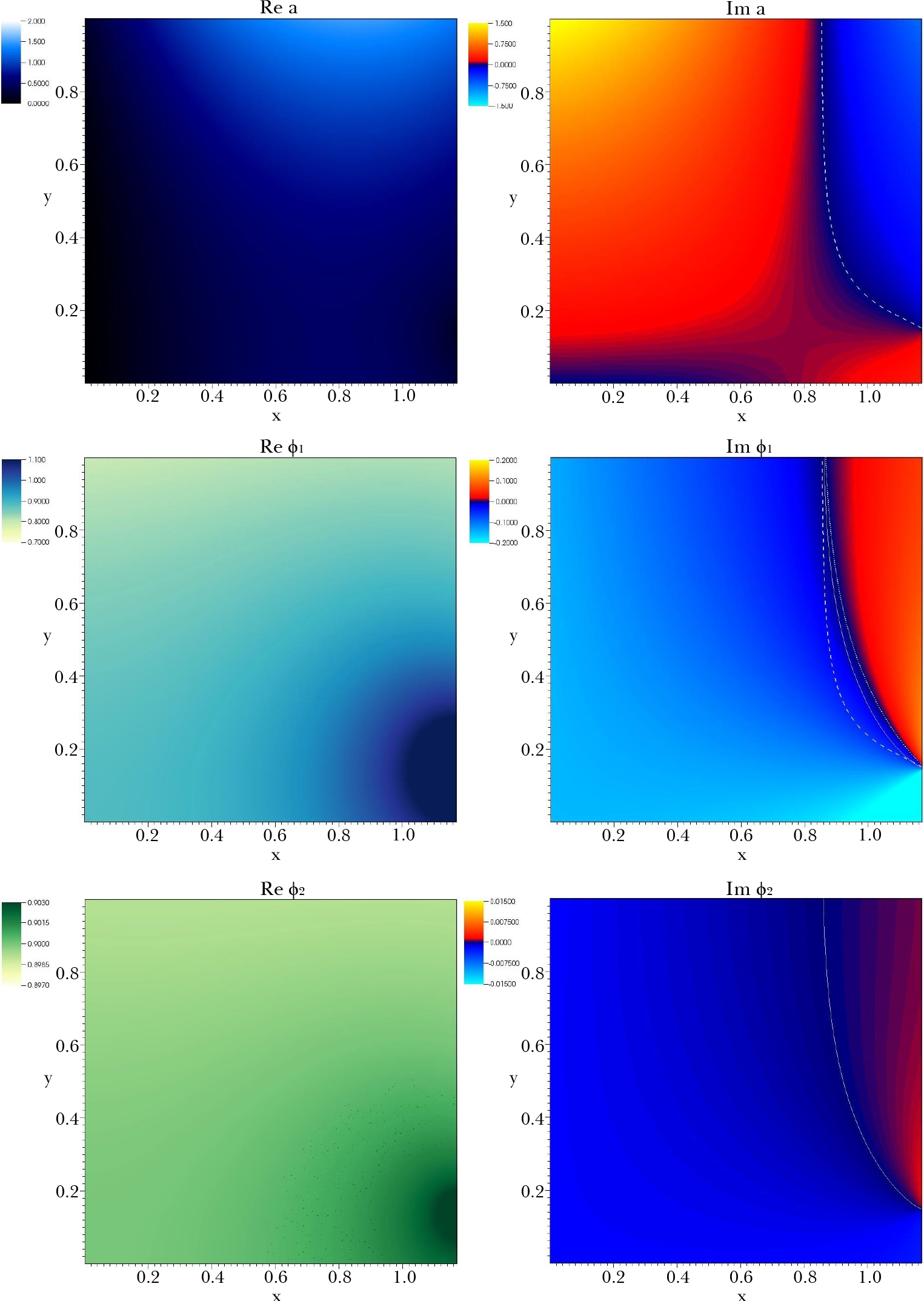}
\caption{\label{fig:classicalized}$a_{r}$, $a_{i}$, $\phi_{1r}$, $\phi_{1i}$, $\phi_{2r}$, and $\phi_{2i}$ over the complex time plane $\tau = x + iy$, for $\mu^{2} = 0.01$, $|\phi_{1}(0)| = |\phi_{2}(0)| = 0.9$ with initial conditions $\theta_{1} \simeq -0.1676$, and $\theta_{2} \simeq -0.0016$. Dashed, dotted, and thin white curves are $a_{i} = 0$, $\phi_{1i}=0$, and $\phi_{2i}=0$, respectively, where we superimposed three curves on the figure of middle-right.}
\end{center}
\end{figure}

\subsection{Behavior of non-classicalized field}

\paragraph{Equations of motion} Equations of motion in Lorentzian signatures are separated by real parts
\begin{eqnarray}
\label{eq:re1}0 &=& \ddot{a}_{r}+ \frac{8\pi}{3} a_{r} \left( \dot{\phi}^{2}_{1r} + \dot{\phi}^{2}_{2r} - \dot{\phi}^{2}_{1i} - \dot{\phi}^{2}_{2i} - \frac{1}{2} \left( \phi^{2}_{1r} - \phi^{2}_{1i} + \mu^{2} \phi^{2}_{2r}- \mu^{2} \phi^{2}_{2i} \right) \right) \nonumber \\
&& - \frac{8\pi}{3} a_{i} \left( 2 \dot{\phi}_{1r} \dot{\phi}_{1i} + 2 \dot{\phi}_{2r} \dot{\phi}_{2i} - \phi_{1r} \phi_{1i} - \mu^{2} \phi_{2r} \phi_{2i} \right),\\
0 &=& \ddot{\phi}_{1r} + 3 \left( \frac{\dot{a}_{r}a_{r} + \dot{a}_{i}a_{i}}{a_{r}^{2} + a_{i}^{2}} \right) \dot{\phi}_{1r} - 3 \left(\frac{-\dot{a}_{r}a_{i} + \dot{a}_{i}a_{r}}{a_{r}^{2} + a_{i}^{2}}\right) \dot{\phi}_{1i} + \phi_{1r},\\
0 &=& \ddot{\phi}_{2r} + 3 \left( \frac{\dot{a}_{r}a_{r} + \dot{a}_{i}a_{i}}{a_{r}^{2} + a_{i}^{2}} \right) \dot{\phi}_{2r} - 3 \left( \frac{-\dot{a}_{r}a_{i} + \dot{a}_{i}a_{r}}{a_{r}^{2} + a_{i}^{2}}\right) \dot{\phi}_{2i} + \mu^{2} \phi_{2r},
\end{eqnarray}
and imaginary parts
\begin{eqnarray}
0 &=& \ddot{a}_{i} + \frac{8\pi}{3} a_{r} \left( 2 \dot{\phi}_{1r} \dot{\phi}_{1i} + 2 \dot{\phi}_{2r} \dot{\phi}_{2i} - \phi_{1r}\phi_{1i} - \mu^{2} \phi_{2r}\phi_{2i} \right) \nonumber \\
&& + \frac{8\pi}{3} a_{i} \left( \dot{\phi}^{2}_{1r} + \dot{\phi}^{2}_{2r} - \dot{\phi}^{2}_{1i} - \dot{\phi}^{2}_{2i} - \frac{1}{2} \left( \phi^{2}_{1r} - \phi^{2}_{1i} + \mu^{2} \phi^{2}_{2r}- \mu^{2} \phi^{2}_{2i} \right)  \right),\\
0 &=& \ddot{\phi}_{1i} + 3 \left( \frac{\dot{a}_{r} a_{r} + \dot{a}_{i} a_{i}}{a_{r}^{2} + a_{i}^{2}} \right) \dot{\phi}_{1i} + 3 \left( \frac{- \dot{a}_{r}a_{i} +\dot{a}_{i} a_{r}}{a_{r}^{2} + a_{i}^{2}} \right) \dot{\phi}_{1r} + \phi_{1i},\\
0 &=& \ddot{\phi}_{2i} + 3 \left( \frac{\dot{a}_{r} a_{r} + \dot{a}_{i} a_{i}}{a_{r}^{2} + a_{i}^{2}} \right) \dot{\phi}_{2i} + 3 \left( \frac{- \dot{a}_{r}a_{i} +\dot{a}_{i} a_{r}}{a_{r}^{2} + a_{i}^{2}} \right) \dot{\phi}_{2r} + \mu^{2} \phi_{2i}~,
\end{eqnarray}
where $\dot{~}$ is with respect to the Lorentzian time.

\begin{figure}
\begin{center}
\includegraphics[scale=0.23]{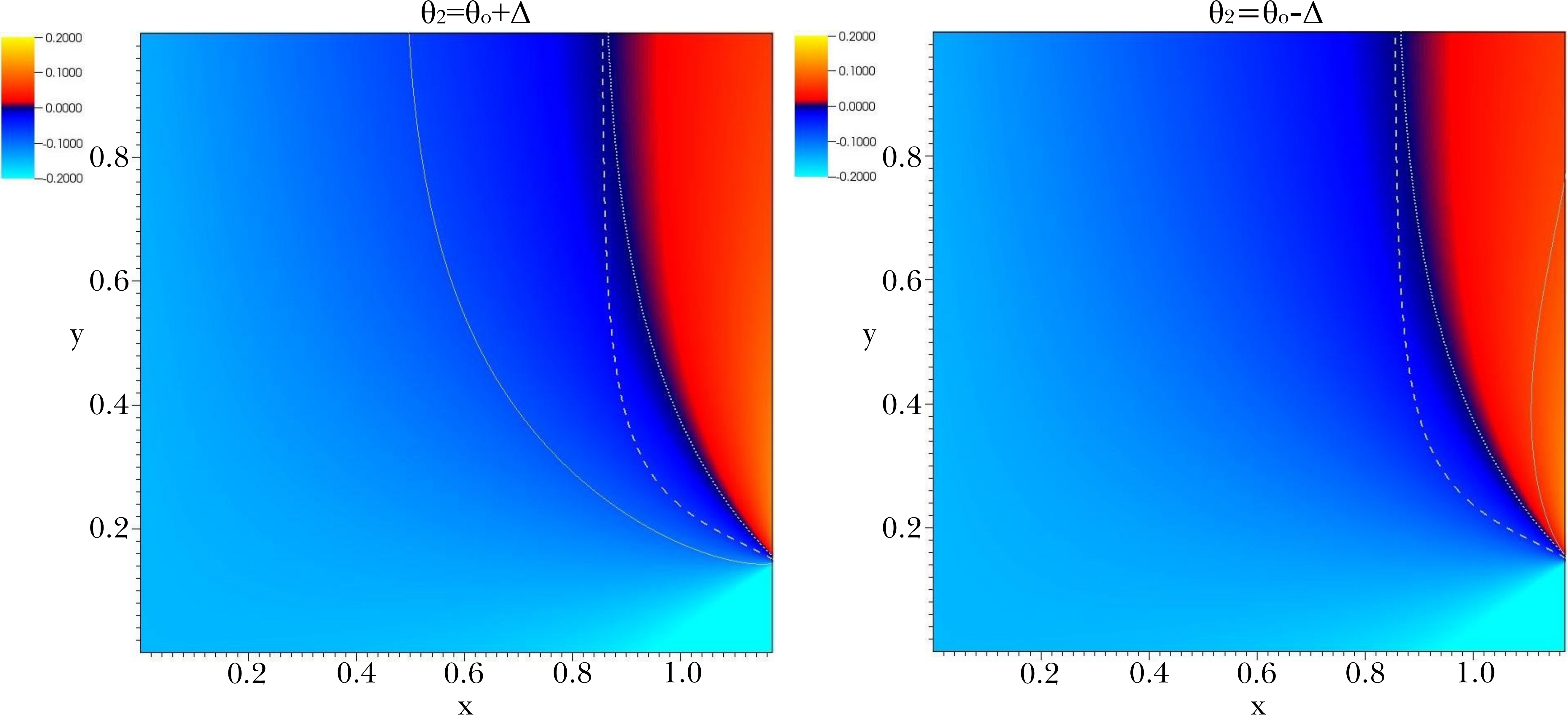}
\caption{\label{fig:tilted}$\phi_{1i}$ for slightly tilted $\theta_{2} = \theta_{o} \pm 2\pi/8192$, where $\theta_{o}$ is the optimized value $\simeq -0.0016$, the plus sign is for left, and the minus sign is for right. Dashed, dotted, and thin white curves are $a_{i} = 0$, $\phi_{1i}=0$, and $\phi_{2i}=0$, respectively.}
\end{center}
\end{figure}

\begin{figure}
\begin{center}
\includegraphics[scale=0.23]{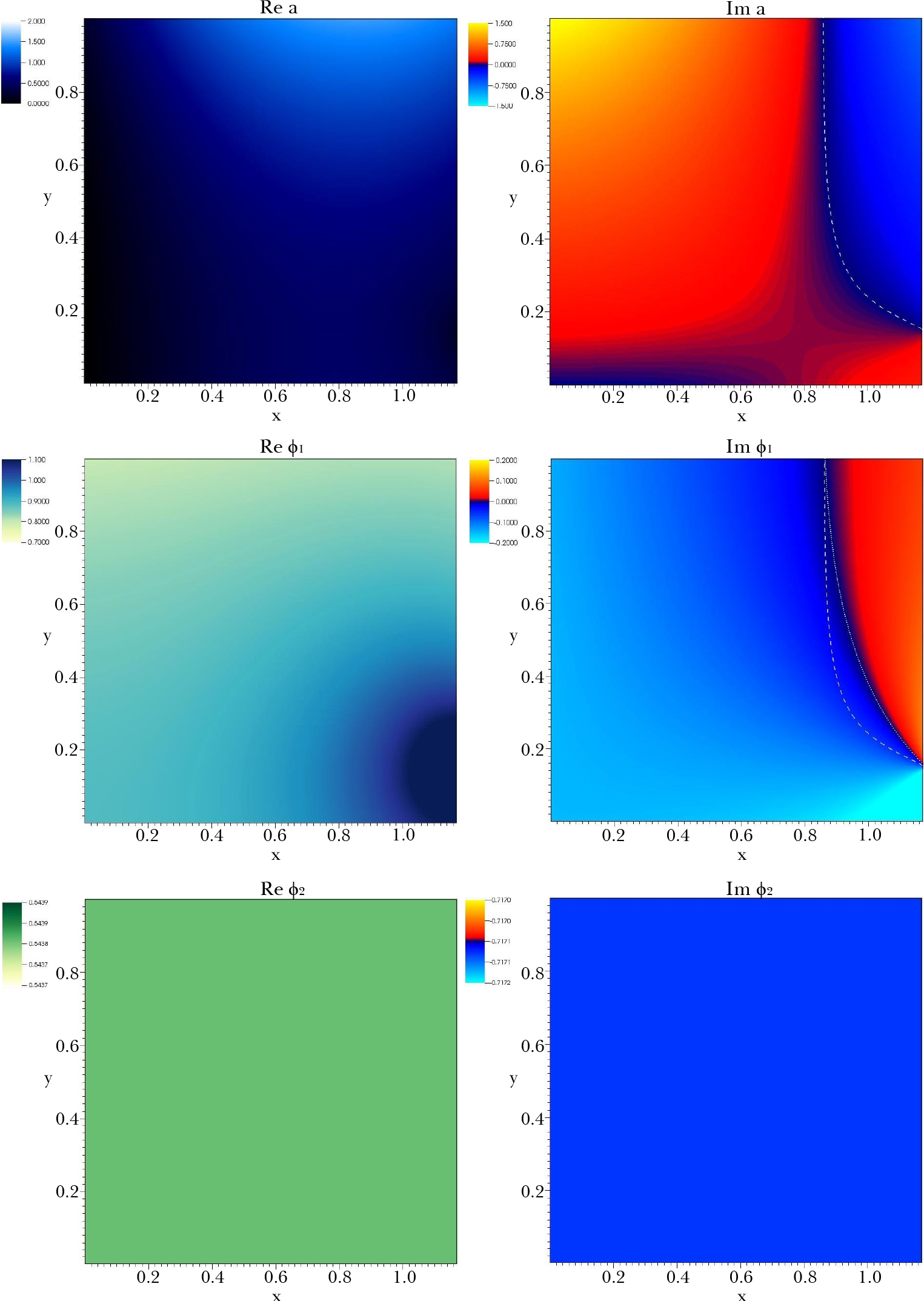}
\caption{\label{fig:figure6}$a_{r}$, $a_{i}$, $\phi_{1r}$, $\phi_{1i}$, $\phi_{2r}$, and $\phi_{2i}$ over the complex time plane $\tau = x + iy$, for $\mu^{2} = 0$, $|\phi_{1}(0)| = |\phi_{2}(0)| = 0.9$ with initial conditions $\theta_{1} \simeq -0.1676$, and $\theta_{2} = \theta_{o} + \Delta_{2}$, where $\Delta_{2} = - 2400\pi /8192$. Dashed and dotted curves are $a_{i} = 0$ and $\phi_{1i}=0$.}
\end{center}
\end{figure}

\paragraph{Existence of $a_{i}, \phi_{1i} \rightarrow 0$ turning time} We first argue that there exists a turning time when $a_{i} \rightarrow 0$ and $\phi_{1i} \rightarrow 0$. As a toy model, let us fix $\mu^{2} = 0.01$, $|\phi_{1}(0)| = |\phi_{2}(0)| = 0.9$. Since $\mu^{2} \ll 1$ and the initial field position is the same, the total energy contribution is dominated by $\phi_{1}$.

If we classicalize two fields at the same time, then the optimized point is $\theta_{1} \simeq -0.1676$, $\theta_{2} \simeq -0.0016$; and along the turning time $X \simeq 0.85$, we obtain the classicalized Lorentzian history. We can solve the same initial condition not only along one time contour, but also over the complex plane (FIG.~\ref{fig:classicalized}) (see also \cite{Battarra:2014xoa}). This result shows that along the turning time $X \simeq 0.85$, three curves (dashed, dotted, and thin white curves, corresponding $a_{i}=0$, $\phi_{1i}=0$, and $\phi_{2i}=0$, respectively) coincide and hence along the Lorentzian time, all fields will be classicalized.

Now let us consider the situation that we tilt $\theta_{2}$ from the optimized value and violates the classicality of $\phi_{2}$. As long as $\left| \Omega_{\phi_{2}} \right|/\left| \Omega_{\phi_{1}} \right| \ll 1$, the effects of $\phi_{2}$ will be very restricted for $a$ and $\phi_{1}$. If the tilted angle increases, then by tuning a proper $\theta_{1}$, again we can obtain a good turning time where $a_{i} \rightarrow 0$ and $\phi_{1i} \rightarrow 0$ are satisfied. For example, in FIG.~\ref{fig:tilted}, we tilt $\theta_{2}$ and check that there still exists a turning time $X$ that satisfies $a_{i}, \phi_{1} \rightarrow 0$.

For more realistic applications, in FIG.~\ref{fig:figure6}, we demonstrated a case when the tilted value is much larger $\theta_{2} = \theta_{o} - 2400\pi /8192$ to demonstrate a phantom phase. In this case, we choose $\mu = 0$ to apply for a realistic cosmological model that should satisfy $\mu \ll 1$ (see details in SEC.~\ref{sec:imp}). Even though the tilted value is larger than the optimized value, still the classicality of $a$ and $\phi_{1}$ is robust.

\paragraph{Embedding in quintom model} FIG.~\ref{fig:tilted} has shown the existence of a history that satisfies $a_{i} \rightarrow 0$ and $\phi_{1i} \rightarrow 0$. We have already demonstrated this numerically. To be prudent, we further check its consistency through analytic approximations. In this regard, if we choose the proper turning time that approximately\footnote{In numerical analysis, $|a_{i}|/|a_{r}|$ and $|\phi_{1i}|/|\phi_{1r}|$ rapidly approaches to zero and hence (although $a_{i}$ and $\phi_{1i}$ are not exactly zero) this is a very good approximation.} satisfies $a_{i} \rightarrow 0$, $\phi_{1i} \rightarrow 0$, and $\dot{a}_{r}/a_{r} \equiv H$, then equations are simplified by
\begin{eqnarray}
\label{eq:01} 0 &=& \dot{H} + H^{2} + \frac{8\pi}{3} \left( \dot{\phi}^{2}_{1r} + \dot{\phi}^{2}_{2r} - \dot{\phi}^{2}_{2i} - \frac{1}{2} \left( \phi^{2}_{1r} + \mu^{2} \phi^{2}_{2r}- \mu^{2} \phi^{2}_{2i} \right) \right), \\
\label{eq:02} 0 &=& \ddot{\phi}_{1r} + 3 H \dot{\phi}_{1r} + \phi_{1r},\\
\label{eq:03} 0 &=& 2 \dot{\phi}_{2r} \dot{\phi}_{2i} - \mu^{2} \phi_{2r}\phi_{2i},\\
\label{eq:04} 0 &=& \ddot{\phi}_{2r} + 3 H \dot{\phi}_{2r} + \mu^{2} \phi_{2r},\\
\label{eq:05} 0 &=& \ddot{\phi}_{2i} + 3 H \dot{\phi}_{2i} + \mu^{2} \phi_{2i}.
\end{eqnarray}
Therefore, except Eq.~(\ref{eq:03}) that is related to $a_{i}$, this system of equations are indistinguishable to the system of a quintessence field and a phantom field.

We already found that there exists a direction that satisfies $a_{i} \rightarrow 0$ and hence Eq.~(\ref{eq:03}) should be consistent in the end. We can further check the consistency. During the inflationary regime, we can approximate $H$ as a slowly varying function. Then the follows are solutions:
\begin{eqnarray}
\phi_{2r} = A_{+} e^{-\alpha_{+} t} + A_{-} e^{-\alpha_{-} t}, \;\;\;\;\; \phi_{2i} = B_{+} e^{-\alpha_{+} t} + B_{-} e^{-\alpha_{-} t},
\end{eqnarray}
where
\begin{eqnarray}
\alpha_{\pm} = \frac{3H \pm \sqrt{9H^{2} - 4 \mu^{2}}}{2}.
\end{eqnarray}
If we insert this to Eq.~(\ref{eq:03}), then this term behaves as
\begin{eqnarray}
\propto e^{-\left(3H - \sqrt{9H^{2}-4\mu^{2}}\right)t}
\end{eqnarray}
and hence as time goes on Eq.~(\ref{eq:03}) will be satisfied. This implies that as time goes on, i.e., as $a_{i}$ and $\phi_{1i}$ decay to zero, Eq.~(\ref{eq:03}) will be automatically satisfied\footnote{We may further choose $H = \mu/2$ to automatically cancel Eq.~(\ref{eq:03}), but we will not further restrict our parameters. Since $|a_{i}/a_{r}| \ll 1$, as we see in Eq.~(\ref{eq:re1}), contributions to energy-momentum tensors will be well approximated by a quintessence field and a phantom field.}.

\begin{figure}
\begin{center}
\includegraphics[scale=0.4]{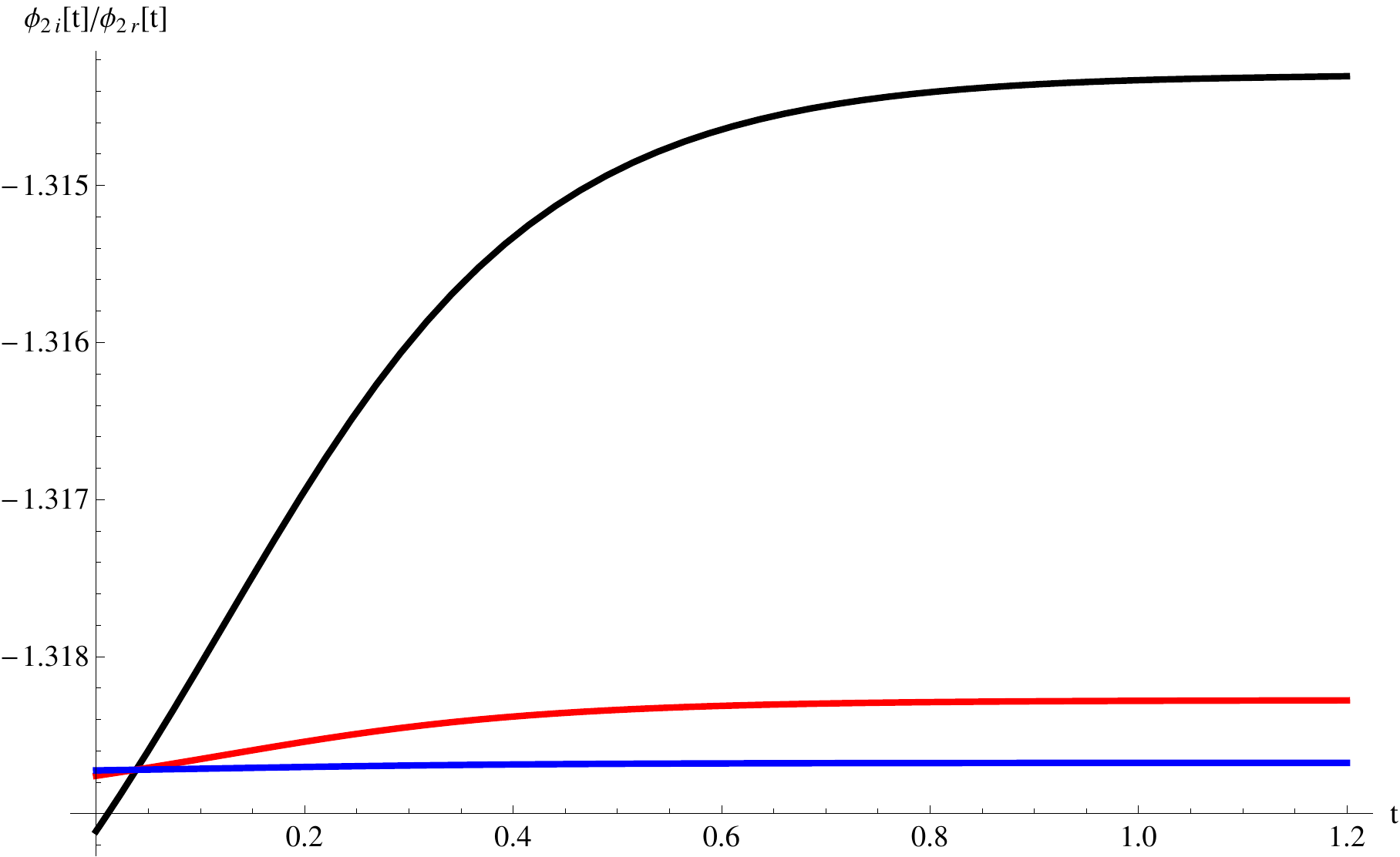}
\caption{\label{fig:comparison}$\phi_{2i}/\phi_{2r}$ for $\theta_{2} = \theta_{o} - 2400\pi/8192$ after the turning time $X$, where $\theta_{o}$ is the optimized value $\simeq -0.0016$. Here, each color denotes different $\mu^{2}$ (black: 0.01, red: 0.001, blue: 0.0001). The ratio approaches a constant as time goes on.}
\end{center}
\end{figure}

\paragraph{Initial conditions} From the above analysis, we thus have various possible initial conditions in the post-inflation period. Let us discuss them in the following:
\begin{itemize}
\item[--] If $A_{+} = B_{-} = 0$, then
\begin{eqnarray}
\frac{|\phi_{2i}|}{|\phi_{2r}|} \propto e^{- (\alpha_{+} - \alpha_{-}) t} = e^{- \sqrt{9H^{2} - 4\mu^{2}}t} \rightarrow 0,
\end{eqnarray}
and hence the classicality of $\phi_{2}$ is satisfied. In other words, the classicality of $\phi_{2}$ is only allowed by a finely-tuned initial condition.
\item[--] If $A_{\pm}$ and $B_{\pm}$ are all non-zero, then
\begin{eqnarray}
\frac{|\phi_{2i}|}{|\phi_{2r}|} \rightarrow \frac{B_{-}}{A_{-}} = \mathrm{const}.
\end{eqnarray}
\item[--] If $A_{-} = B_{+} = 0$, then
\begin{eqnarray}
\frac{|\phi_{2i}|}{|\phi_{2r}|} \propto e^{+ \sqrt{9H^{2} - 4\mu^{2}}t}
\end{eqnarray}
and hence the phantom dominance.
\end{itemize}
If there exists an instanton from the natural parameter space, then $|\phi_{2i}|/|\phi_{2r}| \rightarrow \mathrm{const}$ is the most reasonable condition. Of course, for realistic calculations, $H$ varies with time and hence details are quite complicated. However, as long as we consider the time when $a$ and $\phi_{1}$ are sufficiently classicalized, still this assumption $|\phi_{2i}|/|\phi_{2r}| \rightarrow \mathrm{const}$ is quite reasonable from numerical calculations. In FIG.~\ref{fig:comparison}, we show that the ratio $|\phi_{2i}|/|\phi_{2r}|$ approaches to a constant as time goes on for various choices of $\mu$.

In conclusion, this model is well embedded in a model with a quintessence field $\psi_{q}$ and a phantom field $\psi_{p}$ with the initial conditions satisfying $|\psi_{p}|/|\psi_{q}| \simeq \mathrm{const}$.

\subsection{\label{sec:imp}Generalization: implications to late time cosmology}

\paragraph{Generalization of Hartle-Hawking inspired quintom model} According to the above analysis, for a scalar field system
\begin{eqnarray}
S \supset \int d^4x\sqrt{-g}\left[ -\frac{1}{2} \left(\nabla \phi_{1} \right)^{2} -\frac{1}{2} \left(\nabla \phi_{2} \right)^{2} - V\left(\phi_{1}, \phi_{2}\right)\right], \nonumber
\end{eqnarray}
and after the classicalization of metric and inflaton field $\phi_{1}$, at the end of inflation, it can be transcribed to a two-field model as:
\begin{eqnarray}\label{actionquintom}
S \supset \int d^4x\sqrt{-g}\left[ -\frac{1}{2} \left(\nabla \psi_q \right)^{2} +\frac{1}{2} \left( \nabla \psi_p \right)^{2} - U\left(\psi_p,\psi_q \right)\right],
\end{eqnarray}
where
\begin{eqnarray}
\psi_q &=& \phi_{2r},\\
\psi_p &=& \phi_{2i},\\
U(\psi_q,\psi_p) &=& \mathrm{Re}~V(\phi_{1} = 0, \phi_{2r} + i \phi_{2i}).
\end{eqnarray}
This model, with the signs of the kinetic energy of the two fields being opposite from each other, is actually the ``quintom" model \cite{Feng:2004ad}. One salient property of this model is that its EoS can have a crossing behavior around the cosmological constant boundary $w=-1$, either from above to below or vice versa. There exist varieties of ways to realize quintom behavior (for earliest realizations, see \cite{Guo:2004fq} for two-field models and \cite{Li:2005fm} for single field models with higher derivative). In this paper, we realize a quintom model in a more fundamental way, i.e., from the Hartle-Hawking wave function.

\begin{figure}
\begin{center}
\includegraphics[scale=0.6]{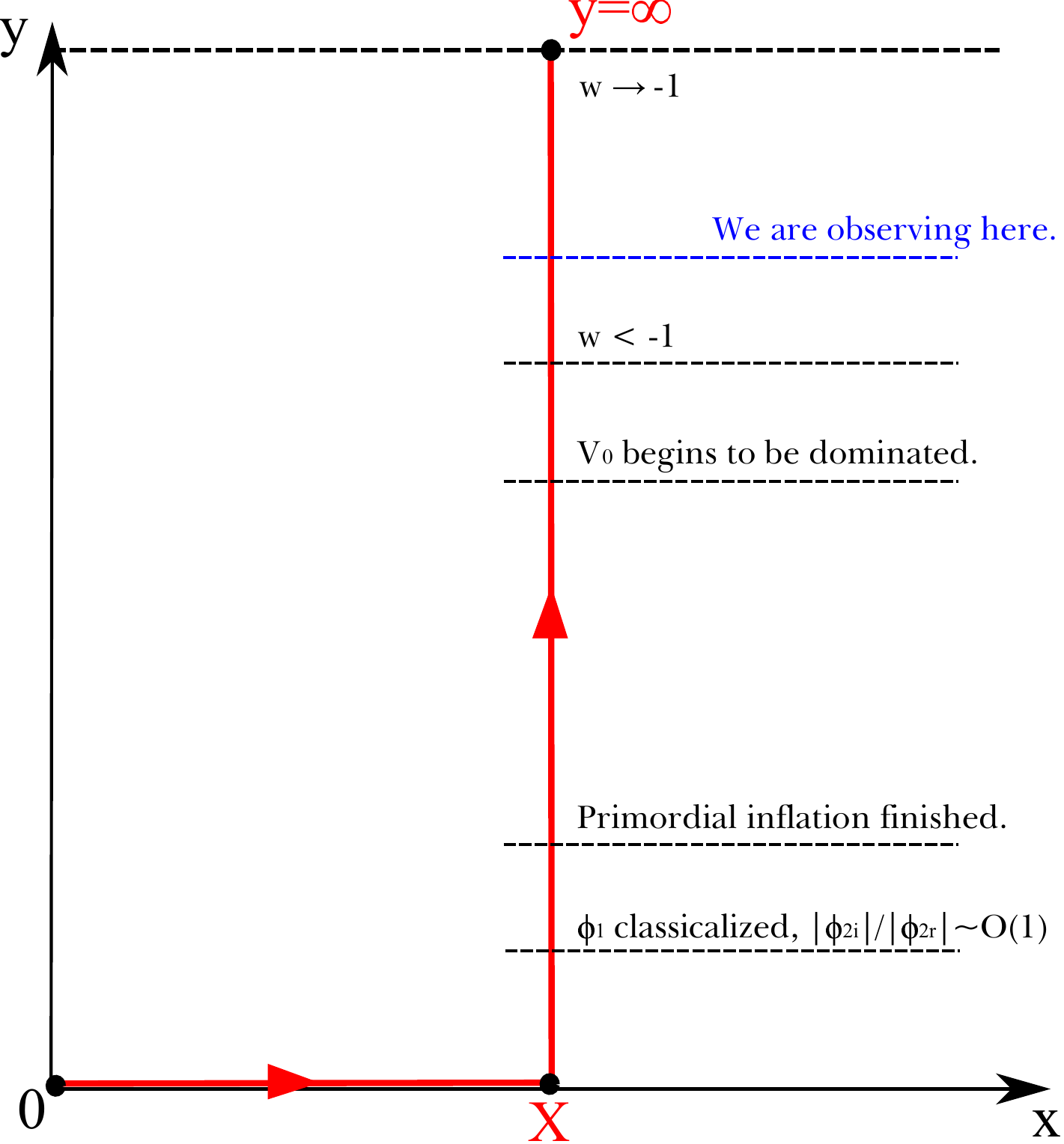}
\caption{\label{fig:path2}A diagram of the conceptual history.}
\end{center}
\end{figure}

\paragraph{Implications to late time cosmology} In our example, we choose the potential to be of quadratic form: $U(\psi_q,\psi_p)=m_2^2\mathrm{Re}[(\phi_{2r}+i\phi_{2i})^2]/2=m_2^2(\psi_q^2-\psi_p^2)/2$. Then according to action (\ref{actionquintom}), the total energy density and pressure of this quintom model are
\begin{eqnarray}
\rho=\frac{1}{2}\dot\psi_q^2-\frac{1}{2}\dot\psi_p^2+\frac{1}{2}m_2^2(\psi_q^2-\psi_p^2), ~~~p=\frac{1}{2}\dot\psi_q^2-\frac{1}{2}\dot\psi_p^2-\frac{1}{2}m_2^2(\psi_q^2-\psi_p^2),
\end{eqnarray}
and the equations of motion for $\psi_q$ and $\psi_p$ are:
\begin{equation}
\ddot\psi_q+3H\dot\psi_q+m_2^2\psi_q=0,~~~\ddot\psi_p+3H\dot\psi_p+m_2^2\psi_p=0,~
\end{equation}
respectively. Thus the equation of state of the whole system is:
\begin{eqnarray}\label{eosquintom}
w = \frac{\dot{\psi}_{q}^{2}-\dot{\psi}_{p}^{2} - m_{2}^{2} \left(\psi_{q}^{2}- \psi_{p}^{2} \right) + 2p_{1} + 2p_{0}}{\dot{\psi}_{q}^{2}-\dot{\psi}_{p}^{2} + m_{2}^{2} \left(\psi_{q}^{2}- \psi_{p}^{2} \right) + 2\rho_{1} + 2\rho_{0}},
\end{eqnarray}
where $p_{1}$ and $\rho_{1}$ are contributions from $\phi_{1}$; $p_{0}$ and $\rho_{0}$ are contributions from the cosmological constant $V_{0}$.

Let us focus on the following points, which has been shown in FIG.~\ref{fig:path2}:
\begin{itemize}
\item[--] During the inflationary phase, the contribution of $V_{0}$ was negligible. However, after inflation ends, $p_{1}$ and $\rho_{1}$ become negligible, while $V_0$ may eventually emerge.
\item[--] In this limit, $\phi_{2}$ remains in the over-damped regime, since we assumed $m_{2}^{2}/V_{0} < 6\pi$. Then
\begin{eqnarray}
\phi_{2r,2i}\propto e^{- \frac{\left(3\tilde{H} - \sqrt{9\tilde{H}^{2}-4\mu^{2}}\right)}{2}t},
\end{eqnarray}
where $\tilde{H}$ is determined by $V_{0}$. We see, once again, that
\begin{eqnarray}
r \equiv \frac{|\psi_{p}|}{|\psi_{q}|}\simeq \mathrm{const}.
\end{eqnarray}
This ratio $r$ will be determined when the field is created by an instanton.
\end{itemize}


In this limit, from Eq.~(\ref{eosquintom}) we have
\begin{eqnarray}
w = \frac{\dot\psi_{q}^{2}-\dot\psi_{p}^{2} - m_{2}^{2}(\psi_{q}^{2}- \psi_{p}^{2} ) - 2V_0}{\dot\psi_{q}^{2}-\dot\psi_{p}^{2} + m_{2}^{2} (\psi_{q}^{2}- \psi_{p}^{2} ) + 2V_0},~~~1+w=\frac{2(\dot\psi_{q}^{2}-\dot\psi_{p}^{2})}{\dot\psi_{q}^{2}-\dot\psi_{p}^{2} + m_{2}^{2} (\psi_{q}^{2}- \psi_{p}^{2} ) + 2V_0}~.
\end{eqnarray}
If we set the initial conditions such that $\dot\psi_{q}^{2}<\dot\psi_{p}^{2}$, then it is natural to have $1+w<0$, i.e., the phantom behavior. However, along with the evolution, the field energy density will eventually become negligible relative to the constant term $V_0$, and the EoS will approach the cosmological constant boundary (CCB) $w=-1$. To see this, it is useful to define the energy density and pressure for each field component as:
\begin{eqnarray}
\rho_q=\frac{1}{2}\dot\psi_q^2+\frac{1}{2}m_2^2\psi_q^2,~~~p_q=\frac{1}{2}\dot\psi_q^2-\frac{1}{2}m_2^2\psi_q^2,~~~
\rho_p=-\frac{1}{2}\dot\psi_p^2-\frac{1}{2}m_2^2\psi_p^2,~~~p_p=-\frac{1}{2}\dot\psi_p^2+\frac{1}{2}m_2^2\psi_p^2,
\end{eqnarray}
such that $\rho_q>0$, $w_q=p_q/\rho_q>-1$, $\rho_p<0$, $w_p=p_p/\rho_p>-1$. Furthermore, from the equations of motion one gets $\rho_q\approx a_r^{-3(1+w_q)}$, $|\rho_p|\approx a_r^{-3(1+w_q)}$, so both $\rho_q$ and the absolute value of $\rho_p$ decrease with time. This means that both $\psi_q$ and $\psi_p$ will have decreasing contribution in the universe, while $V_0$ remains a constant. This is why the universe will eventually be dominated by $V_0$, having $w$ approaching $-1$. However, since the evolution of the two fields are the same except for the initial condition, the relation between $\dot\psi_q$ and $\dot\psi_p$ could be more subtle. If during the evolution it happens that $\dot\psi_p^2$ exceeds $\dot\psi_q^2$, then $w$ will become larger than $-1$, and the quintom behavior will appear.

\begin{figure}
\begin{center}
\includegraphics[scale=0.6]{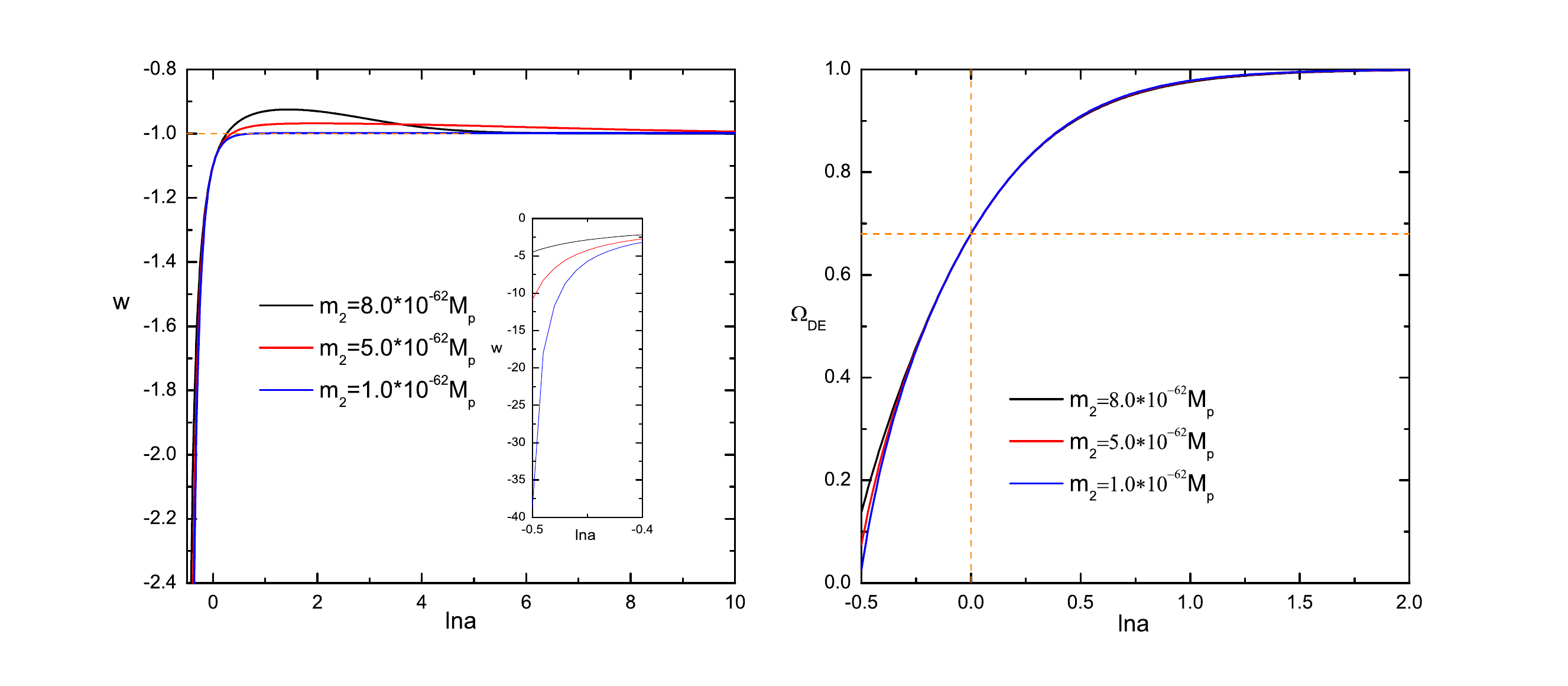}
\caption{\label{fig:all}Left: Evolution of the equation of state $w$ with respect to $\ln a_r$ in our model, where $a_r$ is the classicalized scale factor of our universe. Right: Evolution of the energy density fraction of dark energy $\Omega_{DE}$ with respect to $\ln a_r$ in our model. In the numerical study, we choose $m_2=8.0\times10^{-62}M_p$ (black), $5.0\times10^{-62}M_p$ (red), $1.0\times10^{-62}M_p$ (blue) respectively, while $V_0=0.5\times10^{-123}M_p^4$. Initial conditions: $\psi_{qi}\simeq0.33M_p$ (black), $0.58M_p$ (red), $3.13M_p$ (blue), $\psi_{pi}\simeq-0.005M_p$ (black), $0.055M_p$ (red), $0.68M_p$ (blue), $\dot\psi_{qi}\simeq3.67\times10^{-62}M_p^2$ (black), $2.79\times10^{-62}M_p^2$ (red), $1.55\times10^{-62}M_p^2$ (blue), $\dot\psi_{pi}\simeq4.93\times10^{-62}M_p^2$ (black), $4.80\times10^{-62}M_p^2$ (red), $4.56\times10^{-62}M_p^2$ (blue). }
\end{center}
\end{figure}

In FIG.~\ref{fig:all}, we draw three cases of evolutions in our model. We start from a phantom phase with $w<-1$, with different initial conditions. One can see from the plot that although the initial values are different, they all eventually converge to the $w=-1$ line, which confirms the above analysis. Moreover, two of the three lines display crossing behavior, and the other one approaches $-1$ directly from below. We also plot the evolution of the energy density fraction $\Omega_{DE}$ for the three cases. All of which shows that in the future $\Omega_{DE}\rightarrow-1$, namely the universe will be dominated by dark energy. Actually, all the other components (including $\psi_q$, $\psi_p$, matter, radiation, etc) decays other than the constant term $V_0$, so it is an attractor solution that the universe will always be dominated by $V_0$. Furthermore, our plot shows that at the current time ($\ln a=0$) we have $w\simeq1.1$, $\Omega_{DE}\simeq0.68$, which are well within the newest Planck data, which suggests that $w=-1.54_{-0.50}^{+0.62}$ ($2\sigma$, \textit{Planck}2015 TT+lowP)\footnote{From joint analysis of data, the best fitted value of $w$ could be closer to $-1$, for example $w=-1.006_{-0.091}^{+0.085}$ based on \textit{Planck} power spectra, \textit{Planck} lensing, and external data \cite{Ade:2013zuv}.} and $\Omega_\Lambda=0.686\pm0.020$ ($1\sigma$, \textit{Planck}2013) \cite{Ade:2013zuv}.

One important remark is that this model can also be free from the big rip singularity. According to \cite{Caldwell:2003vq}, when the universe is dominated by the dark energy with $w$, the time scale $\Delta t$ for the universe to be of size $a$ is approximately
\begin{eqnarray}\label{bigriptime}
\begin{array}{lll}
\Delta t &= \frac{2}{3(1+w)H_{0} \sqrt{1 - \Omega_{m0}}}\left(a^{\frac{3(1+w)}{2}}-1\right) & ~~~w\neq-1,\\
& = \frac{1}{H_0\sqrt{1-\Omega_{m0}}}\ln a &~~~w=-1,
\end{array}
\end{eqnarray}
where $H_{0}$ is the current Hubble parameter and $\Omega_{m0}$ is the current density fraction of matter in our universe. A big rip singularity occurs when $a\rightarrow\infty$, which will cause:
\begin{eqnarray}
\left\{ \begin{array}{ll}
\Delta t\rightarrow \infty & \;\;\;\;\;\;w=-1,\\
\Delta t=-\frac{2}{3(1+w)H_{0} \sqrt{1 - \Omega_{m0}}} & \;\;\;\;\;\;w<-1.
\end{array} \right.
\end{eqnarray}
Since in our scenario when dark energy dominates the universe ($\Omega_{DE}\rightarrow 1$), $w$ already always converges to (or larger than) $-1$, and therefore it must correspond to the condition that $\Delta t$ goes to infinity. That is, it is impossible for the big rip singularity to occur in a finite time in the future.

In summary, through an explicit example, we showed that our Hartle-Hawking instanton solution can be applied to late time cosmology, with the light fields behaving as phantom and quintessence fields in the quintom model. Since there exist a future attractor where $w=-1$, the big rip singularity is also avoided. Thus one may say that Hartle-Hawking interpretation of the quantum universe can also provide a theoretical basis for the quintom dark energy models, whose EoS can cross the CCB.

\section{\label{sec:int}Interpretations}

The ground state wave function can be represented by the Euclidean path integral \cite{Hartle:1983ai}
\begin{eqnarray}
\Psi_{0} \left[h_{\mu\nu},\chi \right] = \int \mathcal{D}g_{\mu\nu} \mathcal{D}\phi \; e^{-S_{\mathrm{E}}}. \nonumber
\end{eqnarray}
This Euclidean analytic continuation is the origin to introduce complexified fields. The necessity to introduce complexified fields is very clear from some examples, by comparing calculations using instantons and using quantum field theory in de Sitter space \cite{Hwang:2012mf}. These complexified fields are not a problem in general, since we require the reality at the endpoint of the path integral (e.g., asymptotic future infinity).

However, a problem appears in our universe, since we are not \textit{at the endpoint} but \textit{in the process}. If \textit{we are not seeing the exact endpoint}, then it is allowed to see some effects of the imaginary part of a field, i.e., a ghost-like behavior of a scalar field. Since the instanton approximates this wave function, it already contains quantum contributions. Hence, the instanton and its imaginary part are an \textit{emergent result} of the entire path integral.

Can we find an analog of this phenomenon? Hawking radiation can be an example. Hawking radiation can be interpreted by using a particle propagator \cite{Hartle:1976tp}. The particle propagator can be approximated by a classical path over the Euclidean analytic continuation. This process can be interpreted as follows: a particle comes out from the event horizon backward in time (or oppositely, one can say that a negative energy particle comes into the black hole forward in time) and the same energy particle is detected at the asymptotic future infinity. The classicality will be imposed at the future infinity; but as long as a particle moves backward in time, the bulk description cannot be classical. Now if we cut a Cauchy surface including inside the event horizon, the Cauchy surface includes ghost-like particles. This can be conceptually related to the fact that the renormalized energy-momentum tensor $\langle T_{\mu\nu} \rangle$ can violate the null energy condition around the horizon. Even though the null energy condition is violated, it does not cause a serious instability, since the effects of the negative energy are emergent results from the entire path integral.

Of course, there are some conceptual differences between black hole physics and cosmology. For a black hole case, the renormalized energy-momentum tensor is an averaged result $\langle T_{\mu\nu} \rangle$, not an independent instanton. On the other hand, for a cosmological case, we are in a special universe and hence we should see a special and independent instanton. Can we justify this phenomenon further? We remain this for a future work. However, in conclusion, it seems that if our universe could be phantom-like (i.e., $w < -1$), this Hartle-Hawking inspired quintom model can be a \textit{legal} way to justify phantomness in terms of quantum physics.

\section{\label{sec:con}Conclusion}

In this paper, we investigated the Hartle-Hawking wave function with a two-scalar-field model. This wave function is well approximated by summing over instantons. In general, these instantons will be complexified, but in order to obtain a well-defined probability, one needs to require the classicality of each instanton, i.e., all fields should be realized at infinity. However, as long as we are an observer not at infinity but at a finite time, it is permissible to observe the imprints of the imaginary part of the fields.

In order to embed this possibility to the late time cosmology, we assumed two massive canonical scalar fields ($\phi_{1}$ is an inflaton and $\phi_{2}$ has a slower direction) plus a cosmological constant with some physical conditions imposed: (1) initially the energy contribution of $\phi_{1}$ is dominant over $\phi_{2}$ and (2) after $\phi_{1}$ decays, $\phi_{2}$ still satisfies the over-damped condition. Then during primordial inflation, the scale factor $a$ and the inflaton field $\phi_{1}$ will be realized sufficiently; and as long as the first condition is satisfied, even if $\phi_{2}$ is not realized, the realization of $a$ and $\phi_{1}$ can still be robust.

Then all effects of $\phi_{2}$ can be negligible during inflation; as our universe approaches the dark energy dominated era, however, the non-classical and super-slow-roll scalar field will contribute to the equation of state. If the amplitude of the imaginary part of $\phi_{2}$ is larger than that of the real part of $\phi_{2}$, then $w < -1$ can be attained. However, as time goes on, all real and imaginary parts must decay to zero, and hence the EoS will either cross the cosmological constant boundary $w=-1$ then reduce to it, having a quintom-like behavior, or go to $-1$ directly like phantom models. In either of the two ways, the EoS only stays below $-1$ for a finite time, so there should be no concern about the big rip singularity problem. Thus our model has shown that Hartle-Hawking wave function can be viewed as a theoretical basis and a possible origin of the quintom dark energy models in late time cosmology.

Usually, the phantomness can be easily introduced by a ghost field. However, a ghost field causes perturbative instability, and hence physically disallowed \cite{Carroll:2003st}. In this paper, the imaginary part of a scalar field behaves as a ghost field with negative kinetic energy; but this term came from a non-perturbative effect of the entire wave function. Therefore, we may say that this phantomness can be an emergent effect of quantum gravity.

In this paper, we only restricted to quadratic potential, but in principle it can be generalized to various potentials based on different motivations. In addition, one may apply the same philosophy to investigate other physical phenomena such as black holes. If further investigations can indeed establish the connection between dark energy and the non-classicallized instantons, then this would be the first evidence of effects emergent from quantum gravity.

\newpage

\section*{Acknowledgment}
PC and DY are supported by Taiwan's National Center for Theoretical Sciences (NCTS), Taiwan's Ministry of Science and Technology (MOST), and the Leung Center for Cosmology and Particle Astrophysics (LeCosPA) of National Taiwan University. Part of this work was carried out in Paris while PC was visiting Coll\`ege de France, Paris Diderot University's Astroparticle Physics and Cosmology Center (APC), and \'Ecole Polytechnique during the fall of 2014.

\end{document}